\documentclass[twosided,10pt]{IEEEtran}

\usepackage{authblk}
\usepackage{amsmath}
\usepackage{amssymb}
\usepackage{graphicx}
\usepackage{array}
\usepackage{makecell}
\usepackage{tikz}
\usepackage{pgfplots}	

\usepackage{booktabs}
\usepackage{floatflt}
\usepackage{enumerate}
\usepackage{psfrag}	
\usepackage{array}
\usepackage{multirow,hhline}
\usepackage{exscale}
\usepackage{color}
\usepackage{colortbl}
\usepackage{epsfig,subfigure}
\usepackage{cite}
\usepackage{amsthm}
\usepackage{algorithm}
\usepackage{algpseudocode}
\usepackage[font=small]{caption}
\usepackage[latin1]{inputenc} 
\usepackage[T1]{fontenc} 
\usepackage{enumerate}
\usepackage{siunitx}
\usepackage{nopageno}
\usepackage[left=0.58in, right=0.72in, top=0.7in]{geometry}

\newtheorem{definition}{Definition}

\newtheorem{remark}{Remark}

\newtheorem{proposition}{Proposition}

\usetikzlibrary{shapes,snakes}
\usetikzlibrary{calc}
\usetikzlibrary{patterns}
\usetikzlibrary{decorations.pathmorphing} 

\newcommand{\TxAntenna}[3]{
	\coordinate (a) at (#1,#2);
	\draw[line width=0.25pt,scale=(#3)] (a)--($(a)+(0.2,0)$)--($(a)+(0.2,0.7)$)--
	($(a)+(0.1,0.8)$)--($(a)+(0.3,0.8)$)--($(a)+(0.2,0.7)$);
}

\newcommand{\RxAntenna}[4]{
	\coordinate (a) at (#1,#2);
	\draw[line width=0.25pt,scale=(#3)] (a)--($(a)+(-0.2,0)$)--($(a)+(-0.2,0.7)$)--
	($(a)+(-0.1,0.8)$)--($(a)+(-0.3,0.8)$)--($(a)+(-0.2,0.7)$);
}

\begin{document}

\IEEEoverridecommandlockouts
\title{Robust Transceiver Design for MIMO Decode-and-Forward Full-Duplex Relay}
\author{
\IEEEauthorblockN{Ali Kariminezhad and Aydin Sezgin}\\
\thanks{
A. Kariminezhad and A. Sezgin are with the Institute of Digital Communication Systems, Ruhr-Universit\"at Bochum (RUB), Germany (emails: \{ali.kariminezhad, aydin.sezgin\}@rub.de).
}}

\maketitle
\thispagestyle{empty}
\begin{abstract}
Robust transceiver design against unresolvable system uncertainties is of crucial importance for reliable communication. For instance, full-duplex communication suffers from such uncertainties when canceling the self-interference, since the residual self-interference (RSI) remains uncanceled due to imperfect channel knowledge. We consider a MIMO multi-hop system, where the source, the relay and the destination are equipped with multiple antennas. We allow multi-stream beamforming granted by MIMO technique, without restricting the transmissions to single streaming. The relay can operate in either half-duplex or full-duplex mode, and it changes the mode depending on the RSI strength. Furthermore, the relay is assumed to perform a decode-and-forward (DF) strategy. We investigate a robust transceiver design problem, which maximizes the throughput rate of the worst-case RSI under  RSI channel uncertainty bound constraint. The problem turns out to be a non-convex optimization problem. We propose an efficient algorithm to obtain a local optimal solution iteratively. Eventually, we obtain insights on the optimal antenna allocation at the relay input-frontend and output-frontend, for relay reception and transmission, respectively. Interestingly, with less number of antennas at the source than that at the destination, more number of antennas should be used at the relay input-frontend than the relay output-frontend.
\end{abstract}
\vspace*{-0.5cm}
\section{Introduction}
Optimally relaying the signal from a source to a destination for enhancing the network coverage and improving the throughput rate is an active research area~\cite{Kariminezhad2017}. Furthermore, relaying is the only communication means in disaster scenarios if the direct source-destination link is not available. Exploiting a relay for improving communication throughput rate raises several questions to be answered. For instance, how should the relay process the received signal before dispatching it to the destination? Now, relay can receive a signal from the source, process it and transmit it towards the destination in a successive manner. This type of relaying technique is known as half-duplex relaying. However, while receiving a signal at a certain time instant, a relay can simultaneously transmit the previously received signals. This technique is known as full-duplex relaying~\cite{Bliss2007}. 

As a consequence of transmitting and receiving at a common resource unit, the relay is confronted with self-interference (SI). Note that, full-duplex relaying potentially doubles the throughput rate of the communication compared to the half-duplex counterpart, only if the SI is removed completely at the relay input. By physically isolating the transmitter and receiver frontends of the relay, a significant portion of SI can be reduced~\cite{Sabharwal2014,Shankar2012}. Moreover, analog and/or digital signal processing at the relay input can be utilized to cancel a portion of SI~\cite{Shankar20122, Bliss2012, Eltawil2015,Vogt2018}. This can be realized if the estimate of the SI channel state information (CSI) can be obtained at the relay.

By exploiting multiple antennas at the relay, the throughput rate from the source to destination can be improved~\cite{Fan2007,Mo2012}. By using multiple antennas at the relay provides the feasibility of SI cancellation spatially by beamforming techniques such that the impact of SI can be mitigated~\cite{Riihonen2011,Lioliou2010}. For instance, zero-forcing (ZF) beamforming forces the SI to zero at the relay input, however, it is not an optimal scheme in weak SI regimes if the relay is equipped with a limited number of antennas. In contrast~\cite{Ngo2014,Kariminezhad2017SCC} investigate a relaying setup with massive number of antennas at the relay. Here, they show the optimality of ZF process at the relay with very large antenna array.

Further, exploiting multiple antennas at the source and destination can provide the opportunity for improving the communication throughput rate. In a MIMO multi-hop system, the authors in~\cite{Suraweera2014} investigate a amplify-and-forward (AF) relay, where the precoder at the relay and the decoder at the destination is jointly optimized for maximizing the source-destination throughput rate. However, the authors have assumed a single stream transmission, which is not always optimal. The authors in~\cite{Jeong2017} consider a MIMO decode-and-forward (DF) relaying scheme with energy harvesting demands at the relay provided by the source. These works mainly assume the availability of the SI channel for optimal MIMO pre- and post-processing tasks, where the RSI is simply treated as noise with estimated statistical moments. However, these estimates can not be guaranteed to be valid for all applications and scenarios. Hence, the study of a robust design becomes crucially important.

Robust transceiver design against the worst-case RSI channel provides the worst-case threshold for switching between HD and FD operating modes in hybrid relay systems. The authors in~\cite{Taghizadeh2014} investigate a robust design for multi-user full-duplex relaying with multi-antenna DF relay. In that work, the sources and destinations are equipped with single antennas. Moreover, the authors in~\cite{Cirik2016} investigate a robust transceiver design for multi-user MIMO systems for maximizing the weighted sum-rate of the network.

\textit{Contribution:} We consider a DF relay with multiple antennas at the source, relay and destination. In this system, we allow multi-stream beamforming for throughput rate maximization. The achievable rate of the DF full-duplex relaying is cast as a non-convex optimization problem. We propose an efficient algorithm to solve this problem in polynomial time. Finally, the transmit signal covariances at the source and the relay are designed efficiently to be robust against the worst-case RSI channel.
\vspace*{-0.5cm}
\section{System Model}
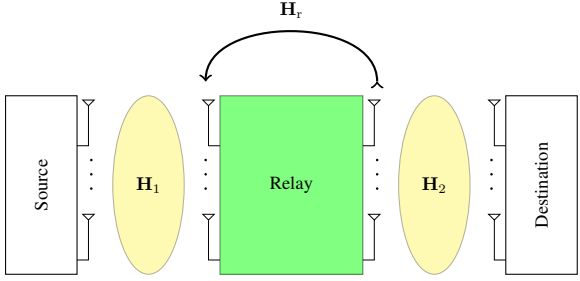
\begin{figure}
\centering
\tikzset{every picture/.style={scale=.95}, every node/.style={scale=0.7}}%
\begin{tikzpicture}
\draw (0,0) rectangle (1,2.5);
\TxAntenna{1}{1.8}{0.8};
\draw (1.2,1.6) circle (0.01cm);
\draw (1.2,1.4) circle (0.01cm);
\draw (1.2,1.2) circle (0.01cm);
\TxAntenna{1}{0.2}{0.8};

\node[rotate=90] at (0.5,1.25){Source};

\draw[fill=green,opacity=0.5] (3,0) rectangle (5,2.5);

\RxAntenna{3}{1.8}{0.8};
\draw (2.8,1.6) circle (0.01cm);
\draw (2.8,1.4) circle (0.01cm);
\draw (2.8,1.2) circle (0.01cm);
\RxAntenna{3}{0.2}{0.8};

\TxAntenna{5}{1.8}{0.8};
\draw (5.2,1.6) circle (0.01cm);
\draw (5.2,1.4) circle (0.01cm);
\draw (5.2,1.2) circle (0.01cm);
\TxAntenna{5}{0.2}{0.8};

\node at (4,1.25){Relay};

\draw (7,0) rectangle (8,2.5);
\RxAntenna{7}{1.8}{0.8};
\draw (6.8,1.6) circle (0.01cm);
\draw (6.8,1.4) circle (0.01cm);
\draw (6.8,1.2) circle (0.01cm);
\RxAntenna{7}{0.2}{0.8};

\node[rotate=90] at (7.5,1.25){Destination};

\draw[fill=yellow,opacity=0.3] (2,1.25) ellipse (0.5cm and 1.25cm);
\node at (2,1.25){$\mathbf{H}_1$};
\draw[fill=yellow,opacity=0.3] (6,1.25) ellipse (0.5cm and 1.25cm);
\node at (6,1.25){$\mathbf{H}_2$};

\draw[thick,->] (5.2,2.7) edge[out=90, in=90] (2.8,2.7);
\node at (4,3.7){$\mathbf{H}_\mathrm{r}$};

\end{tikzpicture}
\vspace*{0.4cm}
\caption{System model of a full-duplex relay}
\label{fig:SystemModel}
\end{figure}

We consider a communication setup from a source equipped with $M$ antennas to a destination with $N$ antennas. The reliable communication is assumed to be only feasible by means of a relay with $K_\mathrm{t}$ transmitter and $K_\mathrm{r}$ receiver antennas at the output and input frontends, respectively. The received signals at the relay and destination are given by
\begin{align}
\mathbf{y}_\mathrm{r}&= \mathbf{H}_1\mathbf{x}_\mathrm{s}+\kappa \mathbf{H}_\mathrm{r}\mathbf{x}_\mathrm{r}+\mathbf{n}_\mathrm{r},\\
\mathbf{y}_\mathrm{d}&= \mathbf{H}_2\mathbf{x}_\mathrm{r}+\mathbf{n}_\mathrm{d},
\end{align}
respectively, where $\kappa\in\{0,1\}$. Notice that, $\kappa=0$ coincides with HD relaying and $\kappa=1$ denotes FD relaying. The transmit signal of the source is denoted by $\mathbf{x}_\mathrm{s}\in\mathbb{C}^{M}$ with the covariance matrix $\mathbf{Q}_\mathrm{s}=\mathbb{E}\{\mathbf{x}_\mathrm{s}\mathbf{x}^H_\mathrm{s}\}$, and the transmit signal of the relay is represented by $\mathbf{x}_\mathrm{r}\in\mathbb{C}^{K_{\mathrm{t}}}$, with the covariance matrix $\mathbf{Q}_\mathrm{r}=\mathbb{E}\{\mathbf{x}_\mathrm{r}\mathbf{x}^H_\mathrm{r}\}$. The additive noise vectors at the relay and destination are denoted by $\mathbf{n}_\mathrm{r}\in\mathbb{C}^{K_\mathrm{r}}$ and $\mathbf{n}_\mathrm{d}\in\mathbb{C}^N$, respectively, which are assumed to follow zero-mean Gaussian distributions with identity covariance matrices. The source-relay channel is represented by $\mathbf{H}_1\in\mathbb{C}^{K_\mathrm{t}\times M}$ and the relay-destination channel is denoted by $\mathbf{H}_2\in\mathbb{C}^{N\times K_\mathrm{r}}$, see~\figurename{ \ref{fig:SystemModel}}. These channels are assumed to be perfectly known. Furthermore, the self-interference (SI) channel at the relay is represented by $\mathbf{H}_\mathrm{r}$, which is assumed to be known only imperfectly. In what follows, we present the achievable throughput rates for the HD and FD relaying. In the next section, we start with the HD relay, where $\kappa=0$.
\vspace*{-0.5cm}
\section{Achievable Rate (Half-Duplex Relay)}
Assuming that the relay applies the decode-and-forward strategy, we consider a simple half-duplex relay where the source and the relay transmit in two subsequent time instances. We formulate the achievable rate between the source and destination nodes as
\begin{align}
R^{\mathrm{HD}}=\frac{1}{2}\min(R^{\mathrm{HD}}_\mathrm{sr},R^{\mathrm{HD}}_\mathrm{rd}),
\end{align}
in which $R^{\mathrm{HD}}_\mathrm{sr}$ and $R^{\mathrm{HD}}_\mathrm{rd}$ are the achievable rates on the source-relay and relay-destination links, respectively. Notice that, in half-duplex relaying the source and relay transmissions are conducted in separate channel uses. These rates are given by
\begin{align}
R^{\mathrm{HD}}_\mathrm{sr}&=\log_2\big|\mathbf{I}_{K_\mathrm{r}}+\mathbf{H}_1\mathbf{Q}_\mathrm{s}\mathbf{H}^H_1\big|,\\
R^{\mathrm{HD}}_\mathrm{rd}&=\log_2\big|\mathbf{I}_{N}+\mathbf{H}_2\mathbf{Q}_\mathrm{r}\mathbf{H}^H_2\big|.
\end{align}
Now, the transmit covariance matrices $\mathbf{Q}_\mathrm{s}\in\mathbb{H}^{M\times M}$ and $\mathbf{Q}_\mathrm{r}\in\mathbb{H}^{K_\mathrm{t}\times K_\mathrm{t}}$ are optimized by maximizing the achievable rate from the source to the destination. Here, the convex cone of hermitian positive semidefinite matrices of dimensions $M\times M$ and $K_\mathrm{t}\times K_\mathrm{t}$ are represented by $\mathbb{H}^{M\times M}$ and $\mathbb{H}^{K_\mathrm{t}\times K_\mathrm{t}}$, respectively. The throughput rate maximization problem is cast as
\begin{subequations}\label{P:HDa}
\begin{align}
\max_{\mathbf{Q}_\mathrm{s},\mathbf{Q}_\mathrm{r}}\quad & \frac{1}{2}\min(R^{\mathrm{HD}}_\mathrm{sr},R^{\mathrm{HD}}_\mathrm{rd})\tag{\ref{P:HDa}}\\
\text{subject to}\quad\quad & \mathrm{Tr}(\mathbf{Q}_\mathrm{s})\leq P_\mathrm{s},\label{P:HDa:ConsA}\\ 
&\mathrm{Tr}(\mathbf{Q}_\mathrm{r})\leq P_\mathrm{r},\label{P:HDa:ConsB}
\end{align}
\end{subequations}
in which the constraints~\eqref{P:HDa:ConsA} and~\eqref{P:HDa:ConsB} are the transmit power constraints and $P_\mathrm{s}$ and $P_\mathrm{r}$ are the transmit power budgets at the source and relay, respectively. Let $\mathbf{Q}_\mathrm{s}=\mathbf{U}_\mathrm{s}\boldsymbol\Gamma_\mathrm{s}\mathbf{U}^H_\mathrm{s}$ and $\mathbf{Q}_\mathrm{r}=\mathbf{U}_\mathrm{r}\boldsymbol\Gamma_\mathrm{r}\mathbf{U}^H_\mathrm{r}$. Since, $R^{\mathrm{HD}}_\mathrm{sr}$ and $R^{\mathrm{HD}}_\mathrm{rd}$ are concave functions of $\mathbf{Q}_\mathrm{s}$ and $\mathbf{Q}_\mathrm{r}$, the solutions are given as~\cite{Telatar99}
\begin{align}
\mathbf{Q}^{\star}_\mathrm{s}={\bf U}^{\star}_\mathrm{s}{\bf\Gamma}^{\star}_\mathrm{s}{\bf U}^{{\star}^H}_\mathrm{s},\ \text{with}\ {\bf U}^\star_\mathrm{s}=\mathbf{R}_1,\label{eq:QsA}\\
\mathbf{Q}^{\star}_\mathrm{r}={\bf U}^{\star}_\mathrm{r}{\bf\Gamma}^{\star}_\mathrm{r}{\bf U}^{{\star}^H}_\mathrm{r},\ \text{with}\ {\bf U}^\star_\mathrm{r}=\mathbf{R}_2.\label{eq:QrA}
\end{align}
Notice that ${\bf R}_1$ and ${\bf R}_2$ correspond to the right singular matrices of ${\bf H}_1$ and ${\bf H}_2$, respectively, with
$
{\bf H}_1={\bf L}_1{\bf \Sigma}_1{\bf R}^H_1,$ and $
{\bf H}_2={\bf L}_2{\bf \Sigma}_2{\bf R}^H_2.
$
The diagonal matrices ${\bf\Gamma}^{\star}_\mathrm{s}$ and ${\bf\Gamma}^{\star}_\mathrm{r}$ are determined by the water-filling algorithm~\cite{Telatar99} as
\begin{align}
{\bf\Gamma}^{\star}_\mathrm{s}&=\left(\tau_\mathrm{s}\mathbf{I}-({\bf\Sigma}^{H}_1{\bf\Sigma}_1)^{-1} \right)^{+},\\
{\bf\Gamma}^{\star}_\mathrm{r}&=\left(\tau_\mathrm{r}\mathbf{I}-({\bf\Sigma}^{H}_2{\bf\Sigma}_2)^{-1} \right)^{+},
\end{align}
respectively. The water levels $\tau_\mathrm{s}$ and $\tau_\mathrm{r}$ are chosen such that they satisfy the power constraint, i.e., $\mathrm{Tr}\left(\tau_\mathrm{s}\mathbf{I}-({\bf\Sigma}_1{\bf\Sigma}^{H}_1)^{-1} \right)=P_\mathrm{s}$, and $\mathrm{Tr}\left(\tau_\mathrm{r}\mathbf{I}-({\bf\Sigma}_2{\bf\Sigma}^{H}_2)^{-1} \right)=P_\mathrm{r}$. Next, we determine the maximum achievable rate for the full-duplex relay.  
\vspace*{-0.2cm}
\section{Achievable Rate (Full-Duplex Relay)}
We assume that an estimate of the self-interference (SI) channel $\mathbf{H}_\mathrm{r}$ is available at the relay denoted by $\tilde{\mathbf{H}}_\mathrm{r}$. Hence, the unknown channel estimation error (residual self-interference channel) represented by $\bar{\mathbf{H}}_\mathrm{r}$ is given as
\begin{align}
\bar{\mathbf{H}}_\mathrm{r}=\mathbf{H}_\mathrm{r}-\tilde{\mathbf{H}}_\mathrm{r}.
\end{align}
In this work, we assume that some portion of the SI is canceled based on the available estimate $\tilde{\mathbf{H}}_\mathrm{r}$, such that only a residual self-interference (RSI) remains. Here, we represent this portion by $\bar{\mathbf{H}}_\mathrm{r}\mathbf{x}_\mathrm{r}$. Considering a full-duplex decode-and-forward relay, the following rate is achievable
\begin{align}
R^{\mathrm{FD}}=\min(R^{\mathrm{FD}}_\mathrm{sr},R^{\mathrm{FD}}_\mathrm{rd}),
\end{align}
in which
\begin{align}
R^{\mathrm{FD}}_\mathrm{sr}&=\log_2\frac{\big|\mathbf{I}_{K_\mathrm{r}}+\mathbf{H}_1\mathbf{Q}_\mathrm{s}\mathbf{H}^H_1+\bar{\mathbf{H}}_\mathrm{r}\mathbf{Q}_\mathrm{r}\bar{\mathbf{H}}^H_\mathrm{r}\big|}{\big|\mathbf{I}_{K_\mathrm{r}}+\bar{\mathbf{H}}_\mathrm{r}\mathbf{Q}_\mathrm{r}\bar{\mathbf{H}}^H_\mathrm{r}\big|},\label{eq:FD_srA}\\
R^{\mathrm{FD}}_\mathrm{rd}&=\log_2\big|\mathbf{I}_{N}+\mathbf{H}_2\mathbf{Q}_\mathrm{r}\mathbf{H}^H_2\big|.
\end{align}
Notice that, with perfect SI channel state information, the SI could be completely removed from the received signal at the relay input-frontend. This doubles the achievable rate correspond to the half-duplex relay. Assuming that a RSI remain uncanceled, a robust transceiver against the worst-case RSI channel is required which is formulated as an optimization problem as follows
\begin{subequations}\label{P:FDa}
\begin{align}
\max_{\mathbf{Q}_\mathrm{s},\mathbf{Q}_\mathrm{r}}\ & \min_{\bar{\mathbf{H}}_\mathrm{r}}\quad  \min(R^{\mathrm{FD}}_\mathrm{sr},R^{\mathrm{FD}}_\mathrm{rd})\tag{\ref{P:FDa}}\\
\text{subject to}\quad\quad\quad & \mathrm{Tr}(\mathbf{Q}_\mathrm{s})\leq P_\mathrm{s},\label{P:FDa:ConsA}\\ 
&\mathrm{Tr}(\mathbf{Q}_\mathrm{r})\leq P_\mathrm{r},\label{P:FDa:ConsB}\\
&\mathrm{Tr}(\bar{\mathbf{H}}_\mathrm{r}\bar{\mathbf{H}}^H_\mathrm{r})\leq T,\label{P:FDa:ConsC}
\end{align}
\end{subequations}
in which the throughput rate of the worst-case RSI channel is maximized. In constraint~\eqref{P:FDa:ConsC}, $T$ represents the RSI channel uncertainty bound. It is important to notice that, without this constraint, the worst-case achievable throughput rate is zero. 
Next, we discuss the optimization problem~\eqref{P:FDa} in details.
\vspace*{-0.5cm} 
\subsection{Robust Transceiver}
We can reformulate~\eqref{eq:FD_srA} as
\begin{align}
R^{\mathrm{FD}}_\mathrm{sr}&=\log_2\big|\mathbf{I}_{K_\mathrm{r}}+\mathbf{H}_1\mathbf{Q}_\mathrm{s}\mathbf{H}^H_1\left(\mathbf{I}+\bar{\mathbf{H}}_\mathrm{r}\mathbf{Q}_\mathrm{r}\bar{\mathbf{H}}^H_\mathrm{r}\right)^{-1}\big|.\label{eq:FD_srB}
\end{align}
Now, by applying the binomial inverse theorem~\cite{Henderson1981}, we arrive at
\begin{align}
R^{\mathrm{FD}}_\mathrm{sr}&=\log_2\big|\mathbf{I}_{K_\mathrm{r}}+\mathbf{H}_1\mathbf{Q}_\mathrm{s}\mathbf{H}^H_1-\nonumber\\
&\mathbf{H}_1\mathbf{Q}_\mathrm{s}\mathbf{H}^H_1\bar{\mathbf{H}}_\mathrm{r}\left(\mathbf{I}_{K_\mathrm{t}}+\mathbf{Q}_\mathrm{r}\bar{\mathbf{H}}^H_\mathrm{r}\bar{\mathbf{H}}_\mathrm{r}\right)^{-1}\mathbf{Q}_\mathrm{r}\bar{\mathbf{H}}^H_\mathrm{r}\big|.\label{eq:FD_srC}
\end{align}
Next, we determine the optimal subspace of the transmit signal from the source and relay.
The RSI channel at the relay is decomposed as
\begin{align}
\bar{\mathbf{H}}_\mathrm{r}&={\bf L}_\mathrm{r}{\bf\Sigma}_\mathrm{r}{\bf R}^{H}_\mathrm{r}.
\end{align}
Interestingly, as given in~\eqref{eq:QsA}, with ${\bf U}_\mathrm{s}={\bf R}_1$ the amount of information extraction at the relay from the source is maximized in a SI-free case. Moreover, with ${\bf U}_\mathrm{r}={\bf R}_2$, the amount of information extraction is maximized at the destination independent of the SI, see~\eqref{eq:QrA}. Notice that, the negative term in the log-determinant expression in~\eqref{eq:FD_srC} is controlled by $\mathbf{Q}_\mathrm{s}$, $\mathbf{Q}_\mathrm{r}$ and $\bar{\mathbf{H}}_\mathrm{r}$. Interestingly, in the log-determinant expression in~\eqref{eq:FD_srC}, the subspace of the negative term can span the subspace of the positive term $\mathbf{H}_1\mathbf{Q}_\mathrm{s}\mathbf{H}^H_1$ by the worst-case RSI channel. This way the worst-case RSI channel can have the most harmful effect on the received signal at the relay.

In what follows, we determine the left and right singular matrices of the worst-case RSI channel, i.e., $\mathbf{L}_\mathrm{r}$ and $\mathbf{R}_\mathrm{r}$. First, we define the maximum number of independent parallel streams that could be supported by the source-relay and relay-destination links. 
\begin{definition}
The degrees-of-freedom $\mathrm{(DoF)}$ supported by the source-relay link and the $\mathrm{DoF}$ of the relay-destination links are defined as
$
\mathrm{DoF}_\mathrm{sr}=\min\{M,K_\mathrm{r}\},$ and 
$\mathrm{DoF}_\mathrm{rd}=\min\{K_\mathrm{t},N\},
$
respectively.
\end{definition}
The following lemma proves useful for the rest of the paper.
\begin{proposition}
Let $\mathrm{DoF}_\mathrm{rd}\geq\mathrm{DoF}_\mathrm{sr}$. Then, the achievable throughput rate from the source to the relay with the worst-case RSI with uncertainty bound $T$ is given by
\begin{subequations}\label{Lemma1}
\begin{align}
\min_{\sigma_{\mathrm{r}_i},\ \forall i}\ &\sum_{i=1}^{\min(M,K_\mathrm{r})} \log_2{\left(1+
\frac{\sigma_{1_i}^2\gamma_{\mathrm{s}_i}}{1+\gamma_{\mathrm{r}_i}\sigma_{\mathrm{r}_i}^2}\right)}\tag{\ref{Lemma1}}\\
\mathrm{s.t.}\quad&\sum_{i=1}^{\min(M,K_\mathrm{r})}\sigma^2_{\mathrm{r}_i}\leq T,
\end{align}
\end{subequations}
where $\sigma_{1_i}$, $\sigma_{\mathrm{r}_i}$, $\gamma_{\mathrm{s}_i}$ and $\gamma_{\mathrm{r}_i}$ are the $i$-th diagonal elements of $\boldsymbol{\Sigma}_1$ ,$\boldsymbol{\Sigma}_\mathrm{r}$, $\boldsymbol{\Gamma}_\mathrm{s}$ ,$\boldsymbol{\Gamma}_\mathrm{r}$, respectively.

\begin{proof}
First, we determine the subspace of the worst-case RSI channel. The left and right singular matrices of the worst-case RSI channel, i.e,  ${\bf L}_\mathrm{r}$ and ${\bf R}_\mathrm{r}$, should project the transmit signal from the relay output on the dimensions spanned by the received signal from the source at the relay input. Let the left singular matrix of the RSI channel be ${\bf L}_\mathrm{r}= {\bf L}_1$. Then, the expression in~\eqref{eq:FD_srC} is formulated as
\begin{align}
\bar{R}^{\mathrm{FD}}_\mathrm{sr}&=\log_2\big|\mathbf{I}_{K_\mathrm{r}}+\mathbf{L}_1\mathbf{\Sigma}_1\mathbf{\Gamma}_\mathrm{s}\mathbf{\Sigma}^{H}_1\mathbf{L}^H_1-\nonumber\\
&\mathbf{L}_1\mathbf{\Sigma}_1\mathbf{\Gamma}_\mathrm{s}\mathbf{\Sigma}^{H}_1\mathbf{\Sigma}_\mathrm{r}{\bf R}^{H}_\mathrm{r}\left(\mathbf{I}_{K_\mathrm{t}}+\mathbf{Q}_\mathrm{r}\bar{\mathbf{H}}^H_\mathrm{r}\bar{\mathbf{H}}_\mathrm{r}\right)^{-1}\mathbf{Q}_\mathrm{r}{\bf R}_\mathrm{r}{\bf\Sigma}^H_\mathrm{r}\mathbf{L}^H_1\big|\nonumber\\
&= \log_2\big|\mathbf{I}_{K_\mathrm{r}}+\mathbf{\Sigma}_1\mathbf{\Gamma}_\mathrm{s}\mathbf{\Sigma}^{H}_1-\nonumber\\
&\mathbf{\Sigma}_1\mathbf{\Gamma}_\mathrm{s}\mathbf{\Sigma}^{H}_1\mathbf{\Sigma}_\mathrm{r}{\bf R}^{H}_\mathrm{r}\left(\mathbf{I}_{K_\mathrm{t}}+\mathbf{Q}_\mathrm{r}\bar{\mathbf{H}}^H_\mathrm{r}\bar{\mathbf{H}}_\mathrm{r}\right)^{-1}\mathbf{Q}_\mathrm{r}{\bf R}_\mathrm{r}{\bf\Sigma}^H_\mathrm{r}\big|.\label{eq:FD_srD}
\end{align}
Notice that the optimal relay transmit covariance matrix $\mathbf{Q}_\mathrm{r}$ lies in the subspace spanned by the left singular matrices of $\mathbf{H}_2$, i.e, $\mathbf{Q}_\mathrm{r}={\bf R}_2{\bf\Gamma}_\mathrm{r}{\bf R}^H_2$, see~\eqref{eq:QrA}.
Now, by ${\bf R}_\mathrm{r}={\bf R}_2$, the negative term in the log-determinant expression spans the subspace of the positive term. Then, the expression in~\eqref{eq:FD_srD} is reformulated as
\begin{align}
\bar{R}^{\mathrm{FD}}_\mathrm{sr}&= \log_2\big|\mathbf{I}_{K_\mathrm{r}}+\mathbf{\Sigma}_1\mathbf{\Gamma}_\mathrm{s}\mathbf{\Sigma}^{H}_1-\nonumber\\
&\mathbf{\Sigma}_1\mathbf{\Gamma}_\mathrm{s}\mathbf{\Sigma}^{H}_1\mathbf{\Sigma}_\mathrm{r}\left(\mathbf{I}_{K_\mathrm{t}}+\mathbf{\Gamma}_\mathrm{r}\mathbf{\Sigma}^H_\mathrm{r}\mathbf{\Sigma}_\mathrm{r}\right)^{-1}\mathbf{\Gamma}_\mathrm{r} {\bf\Sigma}^H_\mathrm{r}\big|\nonumber\\
&=\sum_{i=1} \log_2{(1+\sigma_{1_i}^2\gamma_{\mathrm{s}_i}-\sigma_{1_i}^2\sigma_{\mathrm{r}_i}^2\gamma_{\mathrm{s}_i}\gamma_{\mathrm{r}_i}(1+\gamma_{\mathrm{r}_i}\sigma_{\mathrm{r}_i}^2)^{-1})}\nonumber\\
&=\sum_{i=1}^{\min(M,K_\mathrm{r})} \log_2{\left(1+
\frac{\sigma_{1_i}^2\gamma_{\mathrm{s}_i}}{1+\gamma_{\mathrm{r}_i}\sigma_{\mathrm{r}_i}^2}\right)}.
 \label{eq:FD_srE}
\end{align}
Now, having $\mathrm{DoF}_\mathrm{rd}\geq\mathrm{DoF}_\mathrm{sr}$, the throughput rate over the source-relay link with the worst-case RSI is given by the following optimization problem,
\begin{align}
\min_{\sigma_{\mathrm{r}_i},\ \forall i}\quad \bar{R}^{\mathrm{FD}}_\mathrm{sr}\quad
\mathrm{s.t.}\sum_{i=1}^{\min(M,K_\mathrm{r})}\sigma^2_{\mathrm{r}_i}\leq T,
\end{align}
\end{proof}
\end{proposition}
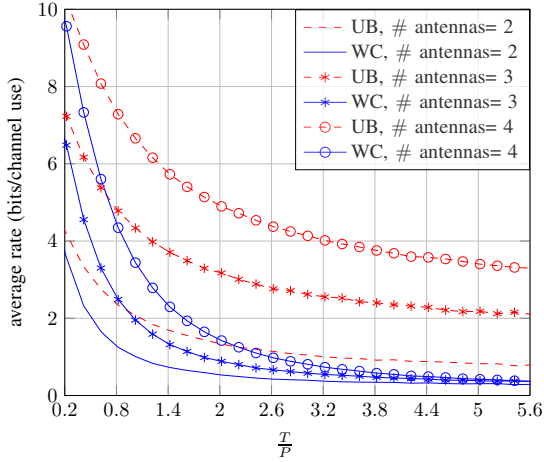
\begin{figure}

\tikzset{every picture/.style={scale=.95}, every node/.style={scale=0.8}}%
\begin{tikzpicture}

\begin{axis}[%
xmin=0.2,
xmax=5.6,
xlabel={$\frac{T}{P}$},
xmajorgrids,
xtick={0.2,0.8,1.4,2,2.6,3.2,3.8,4.4,5,5.6},
ymin=0,
ymax=10,
ylabel={average rate (bits/channel use)},
ylabel near ticks,
ymajorgrids,
ytick={0,2,4,6,8,10,12},
legend style={at={(axis cs: 5.6,10)},anchor=north east,draw=black,fill=white, fill opacity=0.8,legend cell align=left}
]

\addplot [color=red,dashed]
  table[row sep=crcr]{%
  0.02	5.04777554701169\\
  0.22	4.18113500235652\\
  0.42	3.33275841970307\\
  0.62	2.75721533826918\\
  0.82	2.33708915051835\\
  1.02	2.07581251994351\\
  1.22	1.84008439617077\\
  1.42	1.68083568319772\\
  1.62	1.55416044443277\\
  1.82	1.43850640844975\\
  2.02	1.32997437433199\\
  2.22	1.25037291001458\\
  2.42	1.19662225649845\\
  2.62	1.1428528009087\\
  2.82	1.08342417522948\\
  3.02	1.04518271629618\\
  3.22	1.00260800372448\\
  3.42	0.9674666068661\\
  3.62	0.956662526206547\\
  3.82	0.915522577564084\\
  4.02	0.923617125579498\\
  4.22	0.884112114016132\\
  4.42	0.880190681609912\\
  4.62	0.866892081359917\\
  4.82	0.840183568906521\\
  5.02	0.824501124779168\\
  5.22	0.821615039318978\\
  5.42	0.771242505213667\\
  5.62	0.791403222323318\\
  5.82	0.795033578040354\\
  };
\addlegendentry{UB, $\#$ antennas= 2};

\addplot [color=blue]
  table[row sep=crcr]{%
  0.02	5.04006544430001\\
  0.22	3.57325538633924\\
  0.42	2.32167628064635\\
  0.62	1.65583219051744\\
  0.82	1.24659633876128\\
  1.02	1.00931409623609\\
  1.22	0.832586453703777\\
  1.42	0.725458326802039\\
  1.62	0.650085677175754\\
  1.82	0.59226847699902\\
  2.02	0.531005546108135\\
  2.22	0.491647938107233\\
  2.42	0.451829882508842\\
  2.62	0.424549713261466\\
  2.82	0.408914432884257\\
  3.02	0.396828700216316\\
  3.22	0.370188627304943\\
  3.42	0.354875366063222\\
  3.62	0.34663327205656\\
  3.82	0.342984056514327\\
  4.02	0.339765731371898\\
  4.22	0.320111374428858\\
  4.42	0.326359290877647\\
  4.62	0.309695910956522\\
  4.82	0.301564381843428\\
  5.02	0.305685040496176\\
  5.22	0.308534062221038\\
  5.42	0.281278537919998\\
  5.62	0.290185640866138\\
  5.82	0.289435890581432\\
  };
\addlegendentry{WC, $\#$ antennas= 2};

\addplot [color=red,dashed, mark=asterisk,mark options={solid}]
  table[row sep=crcr]{%
  0.02	7.97013522050163\\
  0.22	7.2290736302647\\
  0.42	6.17204413779722\\
  0.62	5.37742837519965\\
  0.82	4.77873450468783\\
  1.02	4.33740417359403\\
  1.22	3.97836643158586\\
  1.42	3.70755195199497\\
  1.62	3.49153009471401\\
  1.82	3.29309264912069\\
  2.02	3.17362855479881\\
  2.22	3.00676715706581\\
  2.42	2.884475379963\\
  2.62	2.76275804970907\\
  2.82	2.71453941594356\\
  3.02	2.61797545477015\\
  3.22	2.55333984018559\\
  3.42	2.52404729790363\\
  3.62	2.42912060054454\\
  3.82	2.39995652293493\\
  4.02	2.349421197352\\
  4.22	2.31646227584094\\
  4.42	2.28251933622496\\
  4.62	2.21147360851337\\
  4.82	2.17243844758227\\
  5.02	2.17895197569643\\
  5.22	2.11854844269771\\
  5.42	2.15444528836961\\
  5.62	2.10197453395566\\
  5.82	2.07461496666072\\
  };
\addlegendentry{UB, $\#$ antennas= 3};

\addplot [color=blue, mark=asterisk,mark options={solid}]
  table[row sep=crcr]{%
  0.02	7.96372496901918\\
  0.22	6.48195399273082\\
  0.42	4.55577852249693\\
  0.62	3.30076775257526\\
  0.82	2.4868178624041\\
  1.02	1.95318098114354\\
  1.22	1.58847204634937\\
  1.42	1.31805460880322\\
  1.62	1.14052421981946\\
  1.82	0.984507574461592\\
  2.02	0.88487252432013\\
  2.22	0.799665424355515\\
  2.42	0.713693171554967\\
  2.62	0.664340245554957\\
  2.82	0.620548209376383\\
  3.02	0.578009500345215\\
  3.22	0.545415010468376\\
  3.42	0.52325198688762\\
  3.62	0.498954668742679\\
  3.82	0.471215912427996\\
  4.02	0.451923939024265\\
  4.22	0.440660294950227\\
  4.42	0.42433082501439\\
  4.62	0.400186344649126\\
  4.82	0.394665973371305\\
  5.02	0.382451814364707\\
  5.22	0.369349645085966\\
  5.42	0.379109831503461\\
  5.62	0.359660203695434\\
  5.82	0.350978768261669\\
  };
\addlegendentry{WC, $\#$ antennas= 3};

\addplot [color=red,dashed, mark=o,mark options={solid}]
  table[row sep=crcr]{%
  0.02	10.891516886589\\
  0.22	10.2244658308264\\
  0.42	9.08691162133928\\
  0.62	8.07726674367587\\
  0.82	7.2844436149728\\
  1.02	6.66054252562779\\
  1.22	6.15741066788434\\
  1.42	5.72805142265364\\
  1.62	5.40213996868812\\
  1.82	5.13779762941802\\
  2.02	4.89648146484921\\
  2.22	4.71927303059782\\
  2.42	4.5363247419769\\
  2.62	4.37186597748191\\
  2.82	4.24694307826173\\
  3.02	4.13470607556557\\
  3.22	4.01701826178147\\
  3.42	3.92285357503233\\
  3.62	3.84669142201825\\
  3.82	3.75576714327859\\
  4.02	3.6825169624085\\
  4.22	3.59635821368025\\
  4.42	3.57661453470616\\
  4.62	3.53492888855263\\
  4.82	3.46928200964652\\
  5.02	3.40169170507052\\
  5.22	3.36071483222529\\
  5.42	3.31642046419543\\
  5.62	3.29551611853844\\
  5.82	3.2710124680662\\
  };
\addlegendentry{UB, $\#$ antennas= 4};

\addplot [color=blue, mark=o,mark options={solid}]
  table[row sep=crcr]{%
  0.02	10.8866765324539\\
  0.22	9.56104900576018\\
  0.42	7.33152835858268\\
  0.62	5.60219107310808\\
  0.82	4.34596179737841\\
  1.02	3.44094883411094\\
  1.22	2.7863828740715\\
  1.42	2.29480200032523\\
  1.62	1.93362596710561\\
  1.82	1.64927230866991\\
  2.02	1.42254035954191\\
  2.22	1.24948656280685\\
  2.42	1.09987604771192\\
  2.62	0.97706975850964\\
  2.82	0.886200228340578\\
  3.02	0.808672462770213\\
  3.22	0.732217783701574\\
  3.42	0.678563620820444\\
  3.62	0.632878172862525\\
  3.82	0.579600776441533\\
  4.02	0.54808151948357\\
  4.22	0.508162970870864\\
  4.42	0.489579000326779\\
  4.62	0.466875613656312\\
  4.82	0.435680921041006\\
  5.02	0.420846902316693\\
  5.22	0.403412581724348\\
  5.42	0.383448915445117\\
  5.62	0.369254833066567\\
  5.82	0.358240840766658\\
  };
\addlegendentry{WC, $\#$ antennas= 4};
  
\end{axis}
\end{tikzpicture}%
\caption{Comparison between the average worst-case achievable rate (WC) and the upperbound (UB). Hypothetically, the singular values of the RSI channel is given. We consider equal number of antennas at all transmitters and receivers. Solid curves: worst-case achievable rates, dashed curves: upperbounds.}
\label{fig:WorstCaseA}
\end{figure}
This shows that, for the worst-case SI channel, the achievable rate of the source-relay link is the sum of achievable rates of $\min(M,K_\mathrm{r})$ data-streams. Notice that by ${\bf L}_\mathrm{r}={\bf L}_1$ and ${\bf R}_\mathrm{r}={\bf R}_2$, the singular directions of the worst-case RSI channel align along the singular direction of the source-relay link. However, this is the worst-case RSI, only if $\mathrm{DoF}_\mathrm{rd}\geq\mathrm{DoF}_\mathrm{sr}$. Otherwise, the singular direction of the worst-case RSI should not align along the singular directions of the source-relay link. Rather, they should lay on the subspace spanned by all singular directions of the source-relay link. Hypothetically, given the singular values of the RSI channel, the comparison between the worst-case achievable rate and the upperbound is depicted in~\figurename{\ref{fig:WorstCaseA}} as a function of $\frac{T}{P}$ for $P=P_\mathrm{s}=P_\mathrm{r}=5$. Notice that, given the singular values of the RSI channel, the worst-case rate is a function of the worst-case singular directions of the RSI channel, which are ${\bf L}_\mathrm{r}={\bf L}_1$ and ${\bf R}_\mathrm{r}={\bf R}_2$. Furthermore, note that, the rates upperbound are for complete RSI channel knowledge.
\begin{remark}
The function $\frac{1}{1+\gamma_{\mathrm{r}_i}\sigma_{\mathrm{r}_i}^2}$ is a monotonically decaying function in both $\gamma_{\mathrm{r}_i}$ and $\sigma_{\mathrm{r}_i}$. Hence, for either $\sigma_{\mathrm{r}_i}=0,\ \forall i$ or $\gamma_{\mathrm{r}_i}=0,\ \forall i$, the achievable rate of the source-relay link is maximized. However, notice that the case with $\sigma_{\mathrm{r}_i}=0,\ \forall i$, represents zero RSI, hence, it is of our interest. However, with $\gamma_{\mathrm{r}_i}=0,\ \forall i$, the relay-destination link throughput rate is zero, hence, this case results in zero source-destination throughput rate. From~\eqref{eq:FD_srE}, allocating less power to the $i$-th stream of the relay-destination link, i.e., $\gamma_{\mathrm{r}_i},\ \forall i\in\{1,\min(M,K_\mathrm{r})\}$, results in an improved achievable rate of the source-relay link, while assuming the worst-case SI channel.
\end{remark}
Now, the remaining question is, how much information bits can be reliably transfered from the source to the destination, with the worst-case ${\sigma_{\mathrm{r}_i},\ \forall i}$?  By determining the left- and right singular matrices of $\mathbf{Q}_\mathrm{s}$, $\mathbf{Q}_\mathrm{r}$ and $\mathbf{H}_\mathrm{r}$, the optimization problem~\eqref{P:FDa} becomes equivalent to
\begin{subequations}\label{P:FDb}
\begin{align}
\max_{\boldsymbol{\gamma}_\mathrm{s},\boldsymbol{\gamma}_\mathrm{r}}\quad & \min_{\boldsymbol{\sigma}_{\mathrm{r}}}\quad  \min(\bar{R}^{\mathrm{FD}}_\mathrm{sr},R^{\mathrm{FD}}_\mathrm{rd})\tag{\ref{P:FDb}}\\
\text{subject to}\quad & \|\boldsymbol{\gamma}_\mathrm{s}\|_1\leq P_\mathrm{s},\label{P:FDb:ConsA}\\ 
&\|\boldsymbol{\gamma}_\mathrm{r}\|_1\leq P_\mathrm{r},\label{P:FDb:ConsB}\\
&\|\boldsymbol{\sigma}^2_\mathrm{r}\|_1\leq T,\label{P:FDb:ConsC}
\end{align}
\end{subequations}
in which, $\boldsymbol{\gamma}_\mathrm{s}=[\gamma_{\mathrm{s}_1},\cdots,\gamma_{\mathrm{s}_{\min(M,K_\mathrm{r}})}]$ and $\boldsymbol{\gamma}_\mathrm{r}=[\gamma_{\mathrm{r}_1},\cdots,\gamma_{\mathrm{r}_{\min(N,K_\mathrm{t}})}]$ and $\boldsymbol{\sigma}_\mathrm{r}=[\sigma_{\mathrm{r}_1},\cdots,\sigma_{\mathrm{r}_{\min(K_\mathrm{t},K_\mathrm{r}})}]$. The optimization problem~\eqref{P:FDb} can be reformulated as
\begin{subequations}\label{P:FDc}
\begin{align}
\max_{R,\boldsymbol{\gamma}_\mathrm{s},\boldsymbol{\gamma}_\mathrm{r}}\quad & \min_{\boldsymbol{\sigma}_{\mathrm{r}}}\quad  R\tag{\ref{P:FDc}}\\
\text{s.t.}\quad & R\leq\nonumber\\ &\sum_{i=1}^{\min(M,K_\mathrm{r})} \log_2{\left(1+
\frac{\sigma_{1_i}^2\gamma_{\mathrm{s}_i}}{1+\gamma_{\mathrm{r}_i}\sigma_{\mathrm{r}_i}^2}\right)},\\
& R\leq\sum_{j=1}^{\min(N,K_\mathrm{t})} \log_2{(1+\sigma_{2_j}^2\gamma_{\mathrm{r}_j})},\\ 
&\eqref{P:FDb:ConsA}-\eqref{P:FDb:ConsC},\nonumber
\end{align}
\end{subequations}
For the purpose of simplification for further discussions, let $M<~K_\mathrm{r}=K_\mathrm{t}<N$. Then, the number of independent data-streams supported by the source-relay and relay-destination links are limited to $M$ and $K_\mathrm{t}$, respectively. Then, the vector of singular values of the worst-case SI channel, i.e., $\boldsymbol{\sigma}_\mathrm{r}$, is composed of $M$ non-zero values and $K_\mathrm{t}-M$ zero values. Interestingly, the robust power allocation at the relay maximizes the information rate of the $i$-th stream by maximizing the term $\frac{1}{1+\gamma_{\mathrm{r}_i}\sigma_{\mathrm{r}_i}^2}$, however the worst-case SI channel for the $i$-th stream represents a $\sigma_{\mathrm{r}_i}$ that minimizes the term $\frac{1}{1+\gamma_{\mathrm{r}_i}\sigma_{\mathrm{r}_i}^2}$. Now, define $\bar{\sigma}^2_{1_i}=
\frac{\sigma^2_{1_i}}{1+\gamma_{\mathrm{r}_i}\sigma_{\mathrm{r}_i}^2}$. Then, the optimization problem~\eqref{P:FDc} is reformulated as
\begin{subequations}\label{P:FDd}
\begin{align}
\max_{R,\boldsymbol{\gamma}_\mathrm{s},\boldsymbol{\gamma}_\mathrm{r}}\quad & \min_{\boldsymbol{\sigma}_{\mathrm{r}}}\quad  R\tag{\ref{P:FDd}}\\
\text{s.t.}\quad & R\leq\sum_{i=1}^{M} \log_2{\left(1+\bar{\sigma}_{1_i}^2\gamma_{\mathrm{s}_i}\right)},\label{P:FDd:ConsA}\\
& R\leq\sum_{j=1}^{K_\mathrm{t}} \log_2{(1+\sigma_{2_j}^2\gamma_{\mathrm{r}_j})},\label{P:FDd:ConsB}\\ 
&\eqref{P:FDb:ConsA}-\eqref{P:FDb:ConsC}.\nonumber
\end{align}
\end{subequations}
The objective of this problem is an affine function. Moreover, the constraints~\eqref{P:FDd:ConsB} and~\eqref{P:FDb:ConsA}-\eqref{P:FDb:ConsC} are convex constraints. However, the constraint~\eqref{P:FDd:ConsA} is a non-convex constraint, since the RHS is not necessarily a concave function of the optimization parameters. Hence, the problem is a non-convex optimization problem. Furthermore, notice that the minimum of the objective function w.r.t. $\boldsymbol{\sigma}_{\mathrm{r}}$ is maximized w.r.t. $R,\boldsymbol{\gamma}_\mathrm{s},\boldsymbol{\gamma}_\mathrm{r}$. Next, we propose an efficient algorithm for obtaining a stationary point.
\begin{algorithm}
\caption{Robust Transceiver Design}
\begin{algorithmic}[1]
\State Set outer-iteration index $l=2$
\State Define $R^{(2)}=1$ and $R^{(1)}=0$
\State Define $\bar{P}_\mathrm{r}^{(l)}=P_\mathrm{r}$
\State Define scalar $c\in [0.9,1)$
\While {$|R^{(l)}-R^{(l-1)}|$ large}
\State Determine $\boldsymbol\gamma^{(l)}_\mathrm{r}=[\tau^{(l)}_\mathrm{r}-\frac{1}{\boldsymbol{\sigma}^2_2}]^{+}$, s.t. $\|\boldsymbol\gamma^{(l)}_\mathrm{r}\|_1=\bar{P}_\mathrm{r}^{(l)}$
\State Determine $\bar{\boldsymbol\gamma}^{(l)}_\mathrm{r}=\boldsymbol\gamma^{(l)}_\mathrm{r}(1:\min(M,K_\text{t}))$
\State Set inner-iteration index $q=2$
\State Define $\boldsymbol{\sigma}^{(2)}_\mathrm{r}=\mathbf{1}^T$ and $\boldsymbol{\sigma}^{(1)}_\mathrm{r}=\mathbf{0}^T$
\State Determine $\boldsymbol\gamma^{(q)}_\mathrm{s}=[\tau^{(q)}_\mathrm{s}-\frac{1}{\boldsymbol{\sigma}^2_1}]^{+}$, s.t. $\|\boldsymbol\gamma^{(q)}_\mathrm{s}\|_1=P_\mathrm{s}$
\While {$\|\boldsymbol{\sigma}^{(q)}_\mathrm{r}-\boldsymbol{\sigma}^{(q-1)}_\mathrm{r}\|_2$ large}
\State Define $\mathbf{u}^{(q)}=\frac{{\boldsymbol{\sigma}}^{(q)^2}_1\odot\boldsymbol\gamma^{(q)}_\mathrm{s}}{\bar{\boldsymbol\gamma}^{(l)}_\mathrm{r}}$
\State Obtain $\boldsymbol{\sigma}^{(q)^2}_\mathrm{r}=[\tau^{(q)}_\mathrm{SI}-\frac{1}{\mathbf{u}^{(q)}}]^{+}$, s.t. $\|\boldsymbol{\sigma}^{(q)^2}_\mathrm{r}\|_1=T$
\State Define $\mathbf{v}^{(q)}=\frac{\tilde{\boldsymbol{\sigma}}^{(q)^2}_\mathrm{1}}
{\mathbf{1}+\boldsymbol\gamma^{(l)}_\mathrm{r}\odot\boldsymbol{\sigma}^{(q)^2}_\mathrm{r}}$
\State Obtain $\boldsymbol\gamma^{(q)}_\mathrm{s}=[\tau^{(q)}_\mathrm{s}-\frac{1}{\mathbf{v}^{(q)}}]^{+}$, s.t. $\|\boldsymbol\gamma^{(q)}_\mathrm{s}\|_1=P_\mathrm{s}$

\State $q=q+1$
\EndWhile
\State Define $\bar{\boldsymbol\gamma}^{(l)}_\mathrm{s}=\boldsymbol\gamma^{(q)}_\mathrm{s}$
\State Define $\bar{\mathbf{v}}^{(l)}=\mathbf
v^{(q)}$
\State Calculate $R^{(l)}_\mathrm{sr}=\sum_{i=1}^{M} \log_2{\left(1+\bar{v}_{i}\bar{\gamma}_{\mathrm{s}_i}\right)} $
\State Calculate $R^{(l)}_\mathrm{rd}=\sum_{j=1}^{K_\mathrm{t}} \log_2{\left(1+\sigma^2_{2_i}{\gamma}_{\mathrm{r}_j}\right)} $
\State $l=l+1$
\State Obtain $R^{(l)}=\min(R^{(l)}_\mathrm{sr},R^{(l)}_\mathrm{rd})$
\State Set $\bar{P}^{(l)}_\mathrm{r}=c\bar{P}^{(l-1)}_\mathrm{r}$
\EndWhile 
\end{algorithmic}
\label{alg:MaterialCharacterization}
\end{algorithm}

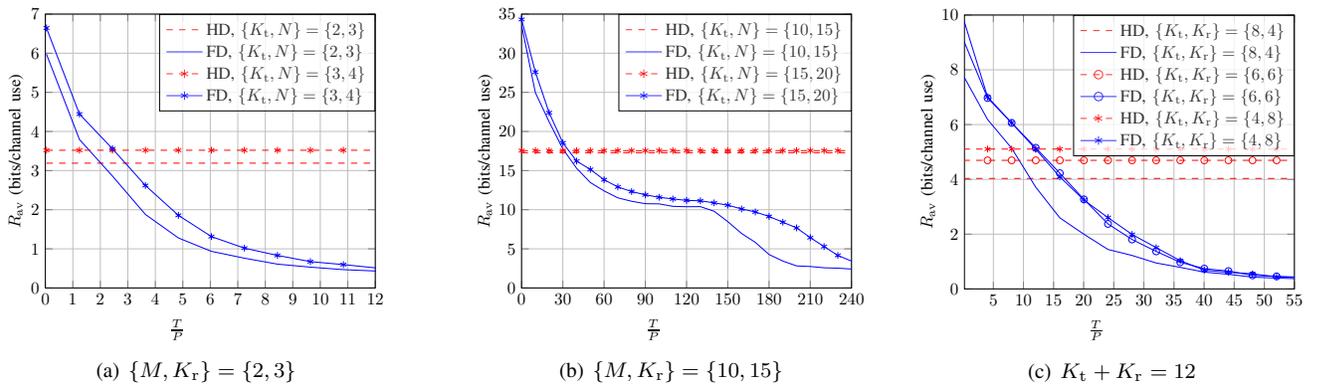
\begin{figure*}
\centering
\begin{minipage}{0.2\textwidth}
\subfigure[$\{M,K_\mathrm{r}\}=\{2,3\}$]{
\tikzset{every picture/.style={scale=.8}, every node/.style={scale=0.8}}%
\begin{tikzpicture}

\begin{axis}[%
xmin=0,
xmax=12,
xlabel={$\frac{T}{P}$},
xmajorgrids,
xtick={0,1,2,3,4,5,6,7,8,9,10,11,12},
ymin=0,
ymax=7,
ylabel={$R_\mathrm{av}$ (bits/channel use)},
ylabel near ticks,
ymajorgrids,
ytick={0,1,2,3,4,5,6,7},
legend style={at={(axis cs: 12,7)},anchor=north east,draw=black,fill=white, fill opacity=0.8,legend cell align=left}
]

\addplot [color=red,dashed]
  table[row sep=crcr]{%
 0.04	3.18973045472121\\
 1.24	3.18973045472121\\
 2.44	3.18973045472121\\
 3.64	3.18973045472121\\
 4.84	3.18973045472121\\
 6.04	3.18973045472121\\
 7.24	3.18973045472121\\
 8.44	3.18973045472121\\
 9.64	3.18973045472121\\
 10.84	3.18973045472121\\
 12.04	3.18973045472121\\
 13.24	3.18973045472121\\
 14.44	3.18973045472121\\
 15.64	3.18973045472121\\
  };
\addlegendentry{HD, $\{K_\mathrm{t},N\}=\{2,3\}$};

\addplot [color=blue]
  table[row sep=crcr]{%
  0.04	5.97793191509576\\
  1.24	3.79510806377925\\
  2.44	2.8613918310694\\
  3.64	1.87853008878625\\
  4.84	1.27800781289388\\
  6.04	0.936220895820177\\
  7.24	0.761015499636601\\
  8.44	0.6096118913737\\
  9.64	0.532263825064486\\
  10.84	0.468661012098091\\
  12.04	0.432718297413609\\
  13.24	0.394952590229019\\
  14.44	0.364817548718091\\
  15.64	0.343424970979551\\
  };
\addlegendentry{FD, $\{K_\mathrm{t},N\}=\{2,3\}$};

\addplot [color=red,dashed, mark=asterisk,mark options={solid}]
  table[row sep=crcr]{%
  0.04	3.51999448364388\\
  1.24	3.51999448364388\\
  2.44	3.51999448364388\\
  3.64	3.51999448364388\\
  4.84	3.51999448364388\\
  6.04	3.51999448364388\\
  7.24	3.51999448364388\\
  8.44	3.51999448364388\\
  9.64	3.51999448364388\\
  10.84	3.51999448364388\\
  12.04	3.51999448364388\\
  13.24	3.51999448364388\\
  14.44	3.51999448364388\\
  15.64	3.51999448364388\\
  };
\addlegendentry{HD, $\{K_\mathrm{t},N\}=\{3,4\}$};

\addplot [color=blue, mark=asterisk,mark options={solid}]
  table[row sep=crcr]{%
  0.04	6.64658530968578\\
  1.24	4.44500195835908\\
  2.44	3.55877935168328\\
  3.64	2.61882056772673\\
  4.84	1.8566513791545\\
  6.04	1.31316261657633\\
  7.24	1.0179646024644\\
  8.44	0.832191629632852\\
  9.64	0.673480980156887\\
  10.84	0.597435813370405\\
  12.04	0.511197681891391\\
  13.24	0.460298830795216\\
  14.44	0.422436693906126\\
  15.64	0.389682844159578\\
  };
\addlegendentry{FD, $\{K_\mathrm{t},N\}=\{3,4\}$};

\end{axis}
\end{tikzpicture}%

\label{fig:HdVsFdB}
}
\end{minipage}
\quad\quad\quad\quad\quad\quad\quad
\begin{minipage}{0.2\textwidth}
\subfigure[$\{M,K_\mathrm{r}\}=\{10,15\}$]{
\tikzset{every picture/.style={scale=.8}, every node/.style={scale=0.8}}%
\begin{tikzpicture}

\begin{axis}[%
xmin=0,
xmax=240,
xlabel={$\frac{T}{P}$},
xmajorgrids,
xtick={0,30,60,90,120,150,180,210,240},
ymin=0,
ymax=35,
ylabel={$R_\mathrm{av}$ (bits/channel use)},
ylabel near ticks,
ymajorgrids,
ytick={0,5,10,15,20,25,30,35},
legend style={at={(axis cs: 240,35)},anchor=north east,draw=black,fill=white, fill opacity=0.8,legend cell align=left}
]

\addplot [color=red,dashed]
  table[row sep=crcr]{%
  0.04	17.2827064366582\\
  10.04	17.2827064366582\\
  20.04	17.2827064366582\\
  30.04	17.2827064366582\\
  40.04	17.2827064366582\\
  50.04	17.2827064366582\\
  60.04	17.2827064366582\\
  70.04	17.2827064366582\\
  80.04	17.2827064366582\\
  90.04	17.2827064366582\\
  100.04	17.2827064366582\\
  110.04	17.2827064366582\\
  120.04	17.2827064366582\\
  130.04	17.2827064366582\\
  140.04	17.2827064366582\\
  150.04	17.2827064366582\\
  160.04	17.2827064366582\\
  170.04	17.2827064366582\\
  180.04	17.2827064366582\\
  190.04	17.2827064366582\\
  200.04	17.2827064366582\\
  210.04	17.2827064366582\\
  220.04	17.2827064366582\\
  230.04	17.2827064366582\\
  240.04	17.2827064366582\\
  250.04	17.2827064366582\\
  260.04	17.2827064366582\\
  270.04	17.2827064366582\\
  280.04	17.2827064366582\\
  290.04	17.2827064366582\\
  };
\addlegendentry{HD, $\{K_\mathrm{t},N\}=\{10,15\}$};

\addplot [color=blue]
  table[row sep=crcr]{%
  0.04	33.5508157815634\\
  10.04	24.9582990023829\\
  20.04	21.2366607786934\\
  30.04	17.7710770263582\\
  40.04	15.2913975075643\\
  50.04	13.4696120661959\\
  60.04	12.4016805402475\\
  70.04	11.5160206666387\\
  80.04	11.0788343091125\\
  90.04	10.7694888346078\\
  100.04	10.7464218522151\\
  110.04	10.4287562061015\\
  120.04	10.3806026995592\\
  130.04	10.3933709790383\\
  140.04	9.80513237034023\\
  150.04	8.47288905022804\\
  160.04	6.96033656071192\\
  170.04	5.82607366888365\\
  180.04	4.27173889242903\\
  190.04	3.44251280014978\\
  200.04	2.81839879678676\\
  210.04	2.74379060914868\\
  220.04	2.57899001234565\\
  230.04	2.53021728213638\\
  240.04	2.41835949619017\\
  250.04	2.55073495797468\\
  260.04	2.57647277065866\\
  270.04	2.50367995188013\\
  280.04	2.564353900746\\
  290.04	2.59998961271894\\
  };
\addlegendentry{FD, $\{K_\mathrm{t},N\}=\{10,15\}$};

\addplot [color=red,dashed, mark=asterisk,mark options={solid}]
  table[row sep=crcr]{%
 0.04	17.5537148917642\\
 10.04	17.5537148917642\\
 20.04	17.5537148917642\\
 30.04	17.5537148917642\\
 40.04	17.5537148917642\\
 50.04	17.5537148917642\\
 60.04	17.5537148917642\\
 70.04	17.5537148917642\\
 80.04	17.5537148917642\\
 90.04	17.5537148917642\\
 100.04	17.5537148917642\\
 110.04	17.5537148917642\\
 120.04	17.5537148917642\\
 130.04	17.5537148917642\\
 140.04	17.5537148917642\\
 150.04	17.5537148917642\\
 160.04	17.5537148917642\\
 170.04	17.5537148917642\\
 180.04	17.5537148917642\\
 190.04	17.5537148917642\\
 200.04	17.5537148917642\\
 210.04	17.5537148917642\\
 220.04	17.5537148917642\\
 230.04	17.5537148917642\\
 240.04	17.5537148917642\\
 250.04	17.5537148917642\\
 260.04	17.5537148917642\\
 270.04	17.5537148917642\\
 280.04	17.5537148917642\\
 290.04	17.5537148917642\\
  };
\addlegendentry{HD, $\{K_\mathrm{t},N\}=\{15,20\}$};

\addplot [color=blue,solid, mark=asterisk,mark options={solid}]
  table[row sep=crcr]{%
  0.04	34.3164472143874\\
  10.04	27.5932251086808\\
  20.04	22.3972070679738\\
  30.04	18.5465700134359\\
  40.04	16.2123489457759\\
  50.04	15.1432011074398\\
  60.04	13.8265613931759\\
  70.04	12.9104925874861\\
  80.04	12.3229776636116\\
  90.04	11.9007829155728\\
  100.04	11.5883694861242\\
  110.04	11.3756503847735\\
  120.04	11.1902600363992\\
  130.04	11.1405447177284\\
  140.04	10.8749506060936\\
  150.04	10.5805956453821\\
  160.04	10.0969023482987\\
  170.04	9.71433461959545\\
  180.04	9.1365340937939\\
  190.04	8.40480845111833\\
  200.04	7.67901969379849\\
  210.04	6.42689423764865\\
  220.04	5.29277356567124\\
  230.04	4.16055073306783\\
  240.04	3.43425259761451\\
  250.04	2.87404221239702\\
  260.04	2.41864542164349\\
  270.04	2.15198127663334\\
  280.04	1.85452068672832\\
  290.04	1.61439314208256\\
  };
\addlegendentry{FD, $\{K_\mathrm{t},N\}=\{15,20\}$};  
  
\end{axis}
\end{tikzpicture}%
\label{fig:HdVsFdD}
}
\end{minipage}
\quad\quad\quad\quad\quad\quad
\begin{minipage}{0.2\textwidth}
\subfigure[$K_\mathrm{t}+K_\mathrm{r}=12$]{
\tikzset{every picture/.style={scale=.8}, every node/.style={scale=0.8}}%
\begin{tikzpicture}

\begin{axis}[%
xmin=0.2,
xmax=55,
xlabel={$\frac{T}{P}$},
xmajorgrids,
xtick={0,5,10,15,20,25,30,35,40,45,50,55},
ymin=0,
ymax=10,
ylabel={$R_\mathrm{av}$ (bits/channel use)},
ylabel near ticks,
ymajorgrids,
ytick={0,2,4,6,8,10},
legend style={at={(axis cs: 55,10)},anchor=north east,draw=black,fill=white, fill opacity=0.8,legend cell align=left}
]

\addplot [color=red,dashed]
  table[row sep=crcr]{%
 0.04	4.03355670119306\\
 4.04	4.03355670119306\\
 8.04	4.03355670119306\\
 12.04	4.03355670119306\\
 16.04	4.03355670119306\\
 20.04	4.03355670119306\\
 24.04	4.03355670119306\\
 28.04	4.03355670119306\\
 32.04	4.03355670119306\\
 36.04	4.03355670119306\\
 40.04	4.03355670119306\\
 44.04	4.03355670119306\\
 48.04	4.03355670119306\\
 52.04	4.03355670119306\\
 56.04	4.03355670119306\\
  };
\addlegendentry{HD, $\{K_\mathrm{t},K_\mathrm{r}\}=\{8,4\}$};

\addplot [color=blue]
  table[row sep=crcr]{%
  0.04	7.77719381959396\\
  4.04	6.18647517587055\\
  8.04	5.1381394653251\\
  12.04	3.71452505378018\\
  16.04	2.59501504325651\\
  20.04	1.99717521189659\\
  24.04	1.44019296431529\\
  28.04	1.22211616187686\\
  32.04	0.947385803760553\\
  36.04	0.787574462236262\\
  40.04	0.613870472037785\\
  44.04	0.535174543781976\\
  48.04	0.430328900052389\\
  52.04	0.396140244396388\\
  56.04	0.36929495131829\\
  };
\addlegendentry{FD, $\{K_\mathrm{t},K_\mathrm{r}\}=\{8,4\}$};

\addplot [color=red,dashed,mark=o, mark options={solid}]
  table[row sep=crcr]{%
  0.04	4.69487160143797\\
  4.04	4.69487160143797\\
  8.04	4.69487160143797\\
  12.04	4.69487160143797\\
  16.04	4.69487160143797\\
  20.04	4.69487160143797\\
  24.04	4.69487160143797\\
  28.04	4.69487160143797\\
  32.04	4.69487160143797\\
  36.04	4.69487160143797\\
  40.04	4.69487160143797\\
  44.04	4.69487160143797\\
  48.04	4.69487160143797\\
  52.04	4.69487160143797\\
  56.04	4.69487160143797\\
  };
\addlegendentry{HD, $\{K_\mathrm{t},K_\mathrm{r}\}=\{6,6\}$};

\addplot [color=blue,mark=o, mark options={solid}]
  table[row sep=crcr]{%
  0.04	9.08683021367302\\
  4.04	6.97117083091761\\
  8.04	6.06259768850812\\
  12.04	5.15280689926556\\
  16.04	4.22580767125593\\
  20.04	3.27068154009634\\
  24.04	2.37398239586416\\
  28.04	1.80433637865438\\
  32.04	1.3702219441655\\
  36.04	0.974140137434874\\
  40.04	0.740327836149415\\
  44.04	0.65202976825448\\
  48.04	0.495991207965511\\
  52.04	0.458111017945286\\
  56.04	0.430017727383298\\
  };
\addlegendentry{FD, $\{K_\mathrm{t},K_\mathrm{r}\}=\{6,6\}$};

\addplot [color=red,dashed,mark=asterisk, mark options={solid}]
  table[row sep=crcr]{%
  0.04	5.10606269606258\\
  4.04	5.10606269606258\\
  8.04	5.10606269606258\\
  12.04	5.10606269606258\\
  16.04	5.10606269606258\\
  20.04	5.10606269606258\\
  24.04	5.10606269606258\\
  28.04	5.10606269606258\\
  32.04	5.10606269606258\\
  36.04	5.10606269606258\\
  40.04	5.10606269606258\\
  44.04	5.10606269606258\\
  48.04	5.10606269606258\\
  52.04	5.10606269606258\\
  56.04	5.10606269606258\\
  };
\addlegendentry{HD, $\{K_\mathrm{t},K_\mathrm{r}\}=\{4,8\}$};

\addplot [color=blue,mark=asterisk, mark options={solid}]
  table[row sep=crcr]{%
  0.04	9.85644482282894\\
  4.04	7.00891698284369\\
  8.04	6.06343949358985\\
  12.04	5.10027710253416\\
  16.04	4.0912204504463\\
  20.04	3.24496409644656\\
  24.04	2.60337488324685\\
  28.04	1.98993573205803\\
  32.04	1.50809046910604\\
  36.04	1.03289439187745\\
  40.04	0.683838482195213\\
  44.04	0.607798682891588\\
  48.04	0.549816836579675\\
  52.04	0.442501542710304\\
  56.04	0.412057172092405\\
  };
\addlegendentry{FD, $\{K_\mathrm{t},K_\mathrm{r}\}=\{4,8\}$};

\end{axis}
\end{tikzpicture}%
\label{fig:HdVsFdE}
}
\end{minipage}

\caption{The transmit power budget at the source and the relay are assumed to be equal, i.e., $P_\mathrm{s}=P_\mathrm
r=P=5$.}
\label{fig:HdVsFdBD}
\end{figure*}

\subsection{Optimization Algorithm}
The proposed algorithm is based on the following intuitions,
\begin{enumerate}
\item given optimal $\bar{\sigma}_{1_i},\ \forall i,$ (genie-aided), the problem is solved by the water-filling algorithm.
\item the rate of the $i$-th stream of the source-relay link is reduced as $\gamma_{\mathrm{r}_i}$ and/or $\sigma_{\mathrm{r}_i}$ increase.
\item at the relay, transmitting with less power than the available power budget, reduces the throughput rate of the relay-destination link, but increases the throughput rate of the source-relay link.
\end{enumerate}

The algorithm is based on successive water-filling procedures with are iterated. This procedure is explained for the $l$-th iteration as follows
\begin{enumerate}[(I)]
\item the relay-destination rate is maximized by allocating more power to better channels, and obtaining a water level $\tau^{(l)}_r$, which satisfies the power budget $P^{(l)}_\mathrm{r}\leq P_\mathrm{r}$,
\item the worst-case SI channel is the one that interferes the strong data-streams (data-streams with high power) received at the relay more than the interference on the weak streams. This is realized by water-filling, and obtaining a water level $\tau^{(l)}_\mathrm{SI}$ that satisfies the water level $T$.
\item having the solutions from (I) and (II), the optimal power allocation at the source is fulfilled by water-filling with water level $\tau^{(l)}_\mathrm{s}$, which satisfies the power constraint $P_\mathrm{s}$,
\item having the solutions from (I), (II) and (III), we compute the achievable source-relay and relay-destination rates, i.e., $R^{(l)}_\mathrm{sr}$ and $R^{(l)}_\mathrm{rd}$. If $|R^{(l)}_\mathrm{rd}-R^{(l)}_\mathrm{sr}|$ is still large, we perform step (I) with a power budget less than $P^{(l)}_\mathrm{r}$.
\end{enumerate}
We present the algorithm pseudo-code in details in Algorithm 1. It is crucial to note that the parameters $\boldsymbol{\sigma}_\mathrm{r}$ and $\boldsymbol{\gamma}_\mathrm{s}$ play a non-collaborative game in the inner-loop for a fixed strategy $\boldsymbol{\gamma}_\mathrm{r}$ which is updated in the outer-loop.

\section{Numerical Results}
We assume equal transmit power budgets at the source and at the relay,  $P=P_\mathrm{s}=P_\mathrm{r}=5$. Moreover, the receiver AWGN variance is assumed to be unity. We investigate the performance of full-duplex relaying with RSI channel uncertainty bound $T$, i.e.,  $\mathrm{Tr}(\bar{\mathbf{H}}_\mathrm{r}\bar{\mathbf{H}}^H_\mathrm{r})\leq T$. We consider the column vectors of the source-relay and the relay-destination channel matrices to be from zero-mean Gaussian distribution with identity covariance matrices. That means, by representing the $i$-th column of $\mathbf{H}_1$ and $j$-th column of $\mathbf{H}_2$ as $\mathbf{h}_{1i}$ and $\mathbf{h}_{2j}$, respectively, we assume $\mathbf{h}_{1i}\sim\mathcal{CN}(\mathbf{0},\mathbf{I})$ and  $\mathbf{h}_{2j}\sim\mathcal{CN}(\mathbf{0},\mathbf{I})$. We perform Monte-Carlo simulations with $L=10^4$ realizations from random channels and noise vectors. Hence, the average worst-case throughput rate is defined as the average of worst-case rates for $L$ randomizations, i.e., 
$
R_{\mathrm{av}}=\frac{1}{L}\sum_{l=1}^{L}R_l.
$
Notice that, for each set of realizations, i.e., $\{\mathbf{H}_1,\mathbf{H}_2,\mathbf{n}_\mathrm{r},\mathbf{n}_\mathrm{d}\}$, we solve the robust transceiver design as is elaborated in Algorithm 1. We run two sets of simulations as described in two following subsections.

\subsection{Antenna Array Increment}
We consider two cases, where the source, relay and destination are equipped with (a)- small antenna array and (b)- large antenna arrays. For these cases, we have
\begin{enumerate}[(a)-]
\item $\{M,K_\mathrm{r}\}=\{2,3\}$ with\\ $\{K_\mathrm{t},N\}=\{2,3\}$ and $\{K_\mathrm{t},N\}=\{3,4\}$
\item $\{M,K_\mathrm{r}\}=\{10,15\}$ with\\ $\{K_\mathrm{t},N\}=\{10,15\}$ and $\{K_\mathrm{t},N\}=\{15,20\}$
\end{enumerate}
These cases are considered to highlight the performance of full-duplex DF relaying as a function of number of antennas with the worst-case RSI. Interestingly, as the number of antennas at the source, relay and destination increase, full-duplex relaying achieves a higher throughput rate even with strong RSI. Furthermore, notice that the worst-case RSI casts strong interference on the strong streams from the source to the destination. With very low RSI power $T\rightarrow 0$, full-duplex almost doubles the throughput rate compared to the half-duplex counterpart. This can be seen in~\figurename{~\ref{fig:HdVsFdBD}}, where the curves cross the vertical axis. However, as $T$ increases, the efficiency of full-duplex operation drops.

Consider the case $\{M,K_\mathrm{r}\}=\{2,3\}$. First let $\{K_\mathrm{t},N\}=\{2,3\}$, where the DoF at the source-relay and relay-destination links are both limited by $2$. In this case, on one hand, the worst-case RSI distributes $T$ over $2$ streams supported by the source-relay link in order to have the most destructive impact. However, on the other hand, the relay transmits with less power, in order to cast less interference on the source-relay link through the RSI channel. Now, with $\{K_\mathrm{t},N\}=\{3,4\}$, the relay-destination link supports $1$ streams more than the source-relay link. Hence, the power controller at the relay will distribute the transmit power over $3$ streams, and only $2$ of those streams cast interference at the relay input (due to the RSI channel). Hence, as can be seen in~\figurename{~\ref{fig:HdVsFdB}}, increasing the $\mathrm{DoF}$ of the relay-destination link does not have significant impact on the achievable rate. This is due to the fact that the source-relay link becomes the communication bottleneck. Similar phenomenon happens with a large antenna array at the source, relay and destination as can be seen in~\figurename{~\ref{fig:HdVsFdD}}

\subsection{Relay Tx/Rx Antenna allocation}
Let the relay have $K_\mathrm{t}+K_\mathrm{r}=12$ in total. Furthermore, suppose that the number of antenna at the source and relay are $\{M,N\}=\{2,10\}$. The question is, from $12$ antennas at the relay, how many should be used for reception for the robust design?. To answer this question, we consider the following scenarios
\begin{enumerate}[(a):]
\item $\{K_\mathrm{t},K_\mathrm{r}\}=\{4,8\}\Rightarrow
\mathrm{DoF}_\mathrm{sr}=2,\ \mathrm{DoF}_\mathrm{rd}=4$
\item $\{K_\mathrm{t},K_\mathrm{r}\}=\{6,6\}\Rightarrow
\mathrm{DoF}_\mathrm{sr}=2,\ \mathrm{DoF}_\mathrm{rd}=6$
\item $\{K_\mathrm{t},K_\mathrm{r}\}=\{8,4\}\Rightarrow
\mathrm{DoF}_\mathrm{sr}=2,\ \mathrm{DoF}_\mathrm{rd}=8$
\end{enumerate}
As can be seen in~\figurename{~\ref{fig:HdVsFdE}}, by using more antennas for reception than for transmission, i.e., $K_\mathrm{r}>K_\mathrm{t}$, at the relay, i.e., case (a), the throughput rate in maximized for both HD relay and worst-case FD relay. This is due to the fact that, increasing the signal-to-noise ratio (SNR) of the source-relay streams enhances the overall throughput rate more than the increase by the $\mathrm{DoF}$ of the relay-destination link. However, notice that in this setup the overall DoF from the source to destination is limited by the DoF of the source-relay link.

\section{Conclusion}
In this paper, we investigated a multi-antenna source communicating with a multi-antenna destination through a multi-antenna relay. The relay is assumed to exploit a decode-and-forward (DF) strategy. The transceivers are designed in order to be robust against the worst-case residual self-interference (RSI). To this end, the worst-case achievable throughput rate is maximized. This optimization problem turns out to be a non-convex problem. Assuming that the degrees-of-freedom (DoF) of the source-relay link is less than the DoF of the relay-destination link, we determined the left and right matrices of the singular vectors of the worst-case RSI channel. Then, the problem is simplified to the optimal power allocation at the transmitters, which guarantees robustness against the worst-case RSI singular values. This simplified problem is still non-convex. Based on the intuitions for optimal power allocation at the source and relay, we proposed an efficient algorithm to capture a stationary point. Hence, in a DF relay with multi-stream beamforming, we determine the critical point where the half-duplex relaying outperforms the full-duplex relaying. This critical point provides a mode-switching threshold in hybrid half-duplex full-duplex relay systems. 

\bibliographystyle{IEEEtran} 
\bibliography{reference}
\end{document}